\documentclass[preprint,12pt]{elsarticle}

\usepackage{graphicx}

\usepackage{amssymb}

\journal{Chem. Phys. Lett.}

\begin{document}

\begin{frontmatter}



\title{On the gradient for metallic systems with a local basis set}


\author{K. Doll$^*$}

\address{Universit\"at Ulm, Institut f\"ur Elektrochemie, 
Albert Einstein-Allee 47, D-89081 Ulm, Germany\\
$^*$ Email: klaus.doll@uni-ulm.de
Phone: ++49-731-50-25441 Fax: ++49-731-50-25409}

\begin{abstract}
The analytical gradient for periodic systems is presented,
for the case of metallic systems. The total energy and the free
energy are computed on
the Hartree-Fock or density functional level, with the wave function
being expanded in terms of Gaussian type orbitals. The expression
for the gradient is similar to the case of insulating systems, when
no thermal broadening is applied. When the occupation of the
states is according to the Fermi function, then the gradient is
consistent with the gradient of the free energy. By comparing
with numerical derivatives, examples demonstrate that a reasonable
accuracy is achieved. 
\end{abstract}


\end{frontmatter}


\section{Introduction}
\label{introduction}
Today, analytical gradients are widely available in electronic structure
codes. In the case of molecules, gradients with respect to the
nuclear position are required, and in solids, in addition, gradients
with respect to the cell parameters. Periodic systems often employ
plane waves as basis functions, but local basis sets are also popular
\cite{CRYSTALbuch,EvarestovBuch}.
Local basis sets, usually atom centered, require the calculation
of derivatives of the basis functions with respect to the nuclear positions,
the Pulay forces \cite{Pulay,Bratoz,Schlegel2000}. 
This holds for the case of
molecular and
periodic \cite{Teramae1983,Teramae1984,Feibelman1991a,Feibelman1991b,HirataIwata1997,Sun1998,Jacquemin1999a,Jacquemin1999b,KudinScuseria2000,Doll2001IJQC,Doll2001CPC,Doll2004TCA,Doll2006TCA,Doll2010,Tobita2003,Weber2006} systems. 
Periodic systems have the
feature that metallic ground states are a possible solution. Metallic
systems are more difficult to treat than insulators, because the position
of the Fermi energy has to be determined, and the integration is only
over a part of the Brillouin zone and thus more difficult than in the 
case of insulators. In the case of Hartree-Fock theory, there are
further problems due to the vanishing density of states at the Fermi
level \cite{AshcroftMermin} and the slow decay of the density matrix
at zero temperature (this is however less problematic at 
finite temperature where the decay is exponential \cite{Goedecker1999}). 
This has motivated the use
of a screened Coulomb operator for the exchange interaction \cite{Heyd2003}.
For an overview of calculations for metals
with Gaussian basis sets, see \cite{Pierobuchartikel}. Some time ago, it
had been argued that the gradient 
requires an extra term due to the shape of the
Fermi surface \cite{Kertesz1984}. 
This will be discussed in the present work, and
it appears that this term is spurious. Numerical 
tests indicate that a reasonable accuracy can be achieved, and the analytical
derivatives agree well with numerical derivatives of the free energy. 

\section{Formalism}
\label{formalism}

\subsection{Zero temperature}
\label{subsection0T}

The analytical gradients for periodic systems, on the Hartree-Fock level,
were introduced by \cite{Teramae1983,Teramae1984}. 
A little later, an article suggested
that an extra term should appear in the case of metals \cite{Kertesz1984},
which will be reconsidered in the following.
A notation similar to
\cite{Teramae1983,Teramae1984,Kertesz1984} is used, for the sake
of simplicity. This corresponds to the case of one dimensional
periodicity, but the argument can analogously be transferred to 
two and three dimensions. The notation is similar to the molecular case
\cite{SzaboOstlund}, apart from the summation over the lattice vectors.

The crystalline orbitals $\Psi_n(\vec r, k)$, with the band index $n$ and
the $k$-point $k$
are expanded in linear combinations of Bloch
functions:

\begin{equation}
\Psi_n(\vec r, k)=\sum_{\mu} c_{\mu n}(k)\psi_{\mu}(\vec r, k)
\end{equation}

with 

\begin{equation}
\psi_{\mu}(\vec r, k)=\frac{1}{\sqrt N}
\sum_j \exp({\rm i}kja)\chi_{\mu}^j(\vec r)
\end{equation}

where $N$ is the number of unit cells in the macro-lattice, or equivalently the
number of reducible $k$-points, and $\chi_{\mu}^j(\vec r)$
being a basis function (e.g. a Gaussian) in cell $j$.
The overlap matrix element 
between orbital $\mu$ in cell 0 and $\nu$ in cell $j$ is obtained as

\begin{equation}
S_{\mu \nu }^{0j}=\int \chi^{0*}_{\mu}(\vec r)\chi_{\nu}^j(\vec r) d^3r
\end{equation}

and its Fourier transform as

\begin{equation}
S_{\mu\nu}(k)=\sum_j\exp({\rm i}kja)S_{\mu \nu}^{0j}
\mbox{ and } S_{\mu \nu}^{0j}=\frac{1}{N}\sum_k S_{\mu\nu}(k)\exp(-{\rm i}kja)
\end{equation}

with the cell
parameter $a$. Because of the orthonormality of the crystalline orbitals,
it holds:

\begin{equation}
\sum_{\mu\nu}c_{\mu m}^*(k)S_{\mu\nu}(k)c_{\nu n}(k)=\delta_{mn}
\end{equation}

The total energy per primitive unit cell
is expressed as in \cite{Teramae1983,Teramae1984,Kertesz1984} as
\begin{equation}
E=\frac{1}{2}\sum_{j,\mu,\nu}(H^{0j}_{\mu\nu}+F^{0j}_{\mu\nu})
P^{j0}_{\nu\mu}+E(NR) 
\end{equation} 
with $H^{0j}_{\mu\nu}$ being the one-electron part of the Fock matrix
element,
$F^{0j}_{\mu\nu}$ the corresponding Fock matrix element:

\begin{eqnarray} & &
F^{0j}_{\mu\nu}=H^{0j}_{\mu\nu}+\sum_{h,l,\tau,\lambda}
P^{lh}_{\lambda\tau}(^{0j}_{\mu\nu}||^{hl}_{\tau\lambda})
\end{eqnarray} 

with $(^{0j}_{\mu\nu}||^{hl}_{\tau\lambda})=(^{hl}_{\tau\lambda}||^{0j}_{\mu\nu})$
being the two-electron integral:

\begin{eqnarray} & &
(^{0j}_{\mu\nu}||^{hl}_{\tau\lambda})=
\int \chi_{\mu}^{0*}(\vec r_1)\chi_{\nu}^j(\vec r_1)
\frac{1}{|\vec r_1-\vec r_2|}
\chi_{\tau}^{h*}(\vec r_2)\chi_{\lambda}^l(\vec r_2)d^3r_1 d^3r_2
\nonumber \\ & & 
  -\frac{1}{2}\int
\chi_{\mu}^{0*}(\vec r_1)\chi_{\lambda}^l(\vec r_1)
\frac{1}{|\vec r_1-\vec r_2|}
\chi_{\tau}^{h*}(\vec r_2)\chi_{\nu}^j(\vec r_2)d^3r_1 d^3r_2
\end{eqnarray}


$P^{j0}_{\nu\mu}$ is the corresponding density matrix element, 
and the nuclear repulsion energy is labelled as $E(NR)$.
Strictly speaking, some of the terms such as $E(NR)$ are divergent for
a periodic system, and a formulation based on e.g. the Ewald and related
methods would be more suitable \cite{VicCoulomb,Vic1994}. 
However, the main issue of the present paper
can easiest be demonstrated with a notation consistent with
references \cite{Teramae1983,Teramae1984,Kertesz1984}, and convergence issues of
the Coulomb sums shall be ignored. 
The Hartree-Fock equations for periodic systems \cite{DelRe,Andre} are:

\begin{eqnarray}
\sum_{\nu}F_{\mu\nu}(k)c_{\nu n}(k)=\sum_{\nu}S_{\mu\nu}(k)c_{\nu n}(k)\epsilon_n(k)
\end{eqnarray}

with $\epsilon_n(k)$ being the eigenvalues.

For metallic systems, the
density matrix is expressed as in \cite{Kertesz1984}:

\begin{eqnarray} 
\label{DensitymatrixT0}
& &
P^{j0}_{\nu\mu}=\frac{2}{N}\sum_{k,n}\exp{({\rm i} kja)}c^*_{\mu n}(k)
c_{\nu n}(k)\theta(E_F-\epsilon_n(k))
\\ & & \nonumber
=\frac{1}{N}\sum_{k} P_{\nu\mu}(k)
\exp({\rm i}kja)
\end{eqnarray}

with the Fermi energy $E_F$ and the
Heaviside function $\theta$. The factor 2 is due to the summation over the
2 spin states.  Due to translational invariance,
relations such as $P^{hl}_{\nu\mu}=P^{h-l \ 0}_{\nu\mu}$ hold.
The derivative of the total energy with respect to a geometrical parameter 
$\frac{\partial E}{\partial X}$
is then obtained as in \cite{Teramae1983,Teramae1984}:

\begin{eqnarray} & & 
\frac{\partial E}{\partial X}=
\sum_{j,\mu,\nu}\frac{\partial H^{0j}_{\mu\nu}}{\partial X}P^{j0}_{\nu\mu} +
\frac{1}{2}\sum_{j,\mu,\nu}\sum_{h,l,\tau,\lambda}
P^{lh}_{\lambda\tau}P^{j0}_{\nu\mu}
\frac{\partial (^{0j}_{\mu\nu}||^{hl}_{\tau\lambda})}{\partial X}
\\ & & \nonumber 
-\sum_{j,\mu,\nu}\frac{\partial S_{\mu\nu}^{0j}}{\partial X}
\sum_{k,n}\frac{2}{N}\exp{({\rm i} kja)}c^*_{\mu n}(k)c_{\nu n}(k)
\theta(E_F-\epsilon_n(k))\epsilon_n(k)
+\frac{\partial E(NR) }{\partial X}
\end{eqnarray}

The expression

\begin{eqnarray}
\sum_{k,n}\frac{2}{N}\exp{({\rm i} kja)}c^*_{\mu n}(k)c_{\nu n}(k)
\theta(E_F-\epsilon_n(k))\epsilon_n(k)
\end{eqnarray}

corresponds to the energy weighted density matrix.

In the following, the derivative of the $\theta$ function shall be considered
in more detail.
When computing the gradient with respect to a geometrical parameter $X$,
then the derivative term $G$ due to the Heaviside function is obtained as

\begin{eqnarray} & &
\label{Ggleichung}
G=\sum_{m,\alpha,\beta}\frac{\partial E}{\partial P_{\beta\alpha}^{m0}}
\frac{\partial P_{\beta\alpha}^{m0}}{\partial \theta(E_F-\epsilon_n(k))}
\frac{\partial \theta(E_F-\epsilon_n(k))}{\partial X}
 \\ & & \nonumber
=\sum_{m,\alpha,\beta}
\frac{1}{2}\left((H^{0m}_{\alpha\beta}+F^{0m}_{\alpha\beta})+\sum_{h,l,\tau,\lambda}
P^{hl}_{\lambda\tau}
(^{0m}_{\alpha\beta}||^{lh}_{\tau\lambda})\right)
\frac{\partial P_{\beta\alpha}^{m0}}{\partial \theta(E_F-\epsilon_n(k))}
\frac{\partial \theta(E_F-\epsilon_n(k))}{\partial X} 
\\ & & \nonumber  
=\sum_{m,\alpha,\beta} F^{0m}_{\alpha\beta}
\frac{\partial P_{\beta\alpha}^{m0}}{\partial \theta(E_F-\epsilon_n(k))}
\frac{\partial \theta(E_F-\epsilon_n(k))}{\partial X} \\ & & \nonumber
=\sum_{m,\alpha,\beta}
\frac{2}{N} F^{0m}_{\alpha\beta} \sum_k\sum_n \exp({\rm i}kma)
c^*_{\alpha n}(k)c_{\beta n}(k)\delta(E_F-\epsilon_n(k))
\left[\frac{\partial E_F}{\partial X}-\frac{\partial \epsilon_n(k)}{\partial
    X}
\right]
 \\ & & \nonumber 
=\frac{2}{N}
\sum_{k,n}\sum_{\alpha,\beta}F_{\alpha\beta}(k)c^*_{\alpha n}(k)c_{\beta n}(k)
\delta(E_F-\epsilon_n(k))
\left[\frac{\partial E_F}{\partial X}-\frac{\partial \epsilon_n(k)}{\partial
    X}\right] \\ & & \nonumber 
=\frac{2}{N}
\sum_{k,n} \sum_{\alpha,\beta}c^*_{\alpha n}(k)S_{\alpha\beta}(k)c_{\beta
  n}(k)\epsilon_n(k)
\delta(E_F-\epsilon_n(k))
\left[\frac{\partial E_F}{\partial X}-\frac{\partial \epsilon_n(k)}{\partial
    X}\right] \\ & & \nonumber  = \frac{2}{N}\sum_{k,n}\epsilon_n(k)
\delta(E_F-\epsilon_n(k))
\left[\frac{\partial E_F}{\partial X}-\frac{\partial \epsilon_n(k)}{\partial
    X}\right]
\\  \nonumber & & 
=\frac{2}{N} \sum_{k,n}E_F\delta(E_F-\epsilon_n(k))
\left[\frac{\partial E_F}{\partial X}-\frac{\partial \epsilon_n(k)}{\partial
X}\right]
= 
\frac{2}{N} \sum_{k,n} E_F \frac{\partial \theta(E_F-\epsilon_n(k))}{\partial X}
\end{eqnarray}

Note that in reference \cite{Kertesz1984}, 
$(H^{0j}_{\alpha\beta}+F^{0j}_{\alpha\beta})$ appears instead of
$2F^{0j}_{\alpha\beta}$, and this appears to be incorrect (see also
the related calculation in \cite{SzaboOstlund}).
With the number of electrons in the unit cell $n_0$,
it follows as in \cite{Kertesz1984}:

\begin{eqnarray} 
\label{TeilchenzahlT0}
& &
n_0
=\sum_{\mu,\nu,j}S^{0j}_{\mu\nu}P^{j0}_{\nu\mu}  \\ & & \nonumber
=\sum_{\mu,\nu,j}S^{0j}_{\mu\nu}
\frac{2}{N}\sum_{k,n}\exp{({\rm i} kja)}c^*_{\mu n}(k)
c_{\nu n}(k)\theta(E_F-\epsilon_n(k))  \\ & & \nonumber
=\frac{2}{N}\sum_{\mu,\nu}\sum_{k,n}S_{\mu\nu}(k)c^*_{\mu n}(k)c_{\nu n}(k)
\theta(E_F-\epsilon_n(k))=\frac{2}{N}\sum_{k,n}\theta(E_F-\epsilon_n(k))
\end{eqnarray}

and, as the particle number is fixed, $\frac{\partial n_0}{\partial X}=0$,
and therefore from equation \ref{Ggleichung}, $G=0$ is obtained:
there is thus no extra term due to the step function, and the same expression
as for the case of insulators \cite{Teramae1983,Teramae1984} can be used for
the derivatives with respect to geometrical parameters.

\subsection{Finite temperature}

An additional problem in the case of metals is the numerical integration
of integrals over the occupied part of the Brillouin zone. This 
problem requires $k$-point meshes as large as possible.
A more efficient way is to apply a finite temperature scheme. 
The calculation
can then be theoretically based on finite temperature
density functional theory \cite{Mermin1965}.
The occupation numbers can be chosen e.g. according to the Fermi function.
Gaussian broadening is another popular scheme 
\cite{FuHo1983,Ho1992,Elsaesser1994}. Further schemes (Lorentzian broadening,
a step function) had been discussed in \cite{Springborg1998}.
The Fermi function has the advantage that the computed free energy has
a direct physical meaning, as it contains the electronic contribution
to the free energy; contributions due to e.g. phonons are however missing
(see, e.g. \cite{Grabowski2007}). The Fermi function is defined as

\begin{equation}
f_{k,n}=\frac{1}{1+\exp((\epsilon_n(k)-E_F)/k_BT)}
\end{equation}

with the Boltzmann constant $k_B$.
A small finite temperature can be introduced, so that the density matrix
becomes 
\begin{eqnarray} & & 
P^{0j}_{\mu\nu}=\frac{2}{N}\sum_{k,n}\exp{(-{\rm i} kja)}c^*_{\nu n}(k)
c_{\mu n}(k)f_{k,n}
\end{eqnarray}

and

\begin{eqnarray} & &
P_{\mu\nu}(k)=2\sum_n c^*_{\nu n}(k) c_{\mu n}(k)f_{k,n}
\end{eqnarray}

Compared to equation \ref{DensitymatrixT0}, the Heaviside function
was replaced with the Fermi function. At zero temperature, the
equations agree. The zero temperature energy
can subsequently be approximated by \cite{Gillan1989}

\begin{equation}
E(0)=\frac{1}{2}((E(T)+F(T))
\end{equation}

with the entropy

\begin{equation}
S(T)=-\frac{2k_B}{N}\sum_{k,n}(f_{k,n} \ln f_{k,n}+(1-f_{k,n})\ln (1-f_{k,n}))
\end{equation}

and the free energy

\begin{equation}
F(T)=E(T)-TS(T)
\end{equation}

$F(T)$ and $E(T)$ are similar at low temperature, and the error should
be relatively small when using $F(T)$ instead of $E(T)$. 
As was pointed out later
\cite{Weinert1992,Wentzcovitch1992,Warren1996},
analytical gradients are, for the case of an occupancy according to the
Fermi function, consistent with the free energy $F(T)$. 
This can be seen by computing the additional terms
due to the entropy:

\begin{eqnarray} & &
\label{entropieterm}
-T\frac{\partial S(T)}{\partial X}=\frac{2k_BT}{N}
\sum_{k,n}\frac{\partial f_{k,n}}
{\partial X} \ln\frac{f_{k,n}}{1-f_{k,n}} \\ & & \nonumber
=-\frac{2}{N}\sum_{k.n}\frac{\partial f_{k,n}}{\partial X}(\epsilon_n(k)-E_F)
=-\frac{2}{N}\sum_{k,n}\frac{\partial f_{k,n}}{\partial X}\epsilon_n(k)
\end{eqnarray}

Here, it was exploited that $\frac{2}{N}\sum_{k,n}f_{k,n}=n_0$ in
analogy to equation \ref{TeilchenzahlT0}
and thus the derivative 
$\frac{2}{N}\sum_{k.n}\frac{\partial f_{k,n}}{\partial X}E_F=0$.
Another term is due to the derivative of the density matrix.

This leads now to an additional term:

\begin{eqnarray} & &
\sum_{j,\alpha,\beta}\frac{\partial E}{\partial P_{\beta\alpha}^{j0}}\sum_{k,n}
\frac{\partial P_{\beta\alpha}^{j0}}{\partial f_{k,n}}
\frac{\partial f_{k,n}}{\partial X} \\ \nonumber & &
=\sum_{j,\alpha,\beta}\frac{\partial E}
{\partial P_{\beta\alpha}^{j0}}
\sum_{k,n}\frac{1}{N}\exp({\rm i}kja)
\frac{\partial P_{\beta\alpha}(k)}{\partial f_{k,n}}\frac{\partial f_{k,n}}
{\partial X} \\ \nonumber & &
=\sum_{j,\alpha,\beta}\frac{2}{N}\sum_{k,n}\exp{({\rm i}
  kja)}F_{\alpha\beta}^{0j}c^*_{\alpha n}(k)
c_{\beta n}(k)\frac{\partial f_{k,n}}{\partial X}
 \\ \nonumber & &
=\frac{2}{N}\sum_{j,\alpha,\beta}c^*_{\alpha n}(k) S_{\alpha\beta}(k)c_{\beta n}(k)
\epsilon_n(k)\frac{\partial f_{k,n}}{\partial X}=\frac{2}{N}
\sum_{k,n}\epsilon_n(k)\frac{\partial f_{k,n}}{\partial X}
\end{eqnarray}

But this term is just equivalent to the entropy term 
in equation \ref{entropieterm}, with opposite sign. As a whole, for 
the derivatives of the free energy with respect to a geometrical parameter
$X$, the two terms containing derivatives of
the occupation number $\frac{\partial f_{k,n}}{\partial X}$ cancel, 
and the expression is:

\begin{eqnarray} & &
\frac{\partial F}{\partial X}=
\frac{\partial E-TS}{\partial X}=
\sum_{j,\mu,\nu}\frac{\partial H^{0j}_{\mu\nu}}{\partial X}P^{j0}_{\nu\mu} +
\frac{1}{2}\sum_{j,\mu,\nu}\sum_{h,l,\tau,\lambda}
P^{lh}_{\lambda\tau}P^{j0}_{\nu\mu}
\frac{\partial (^{0j}_{\mu\nu}||^{hl}_{\tau\lambda})}{\partial X}
\nonumber \\ & &
-\sum_{j,\mu,\nu}\frac{\partial S_{\mu\nu}^{0j}}{\partial X}
\sum_{k,n}\frac{2}{N}\exp{({\rm i} kja)}c^*_{\mu n}(k)c_{\nu n}(k)
f_{k,n}\epsilon_n(k)
+\frac{\partial E(NR) }{\partial X}
\end{eqnarray}

This can be viewed
as a generalization of the result in section 
\ref{subsection0T}, with the $\theta$ function being
replaced with the Fermi function. At zero temperature, this reduces
to the $\theta$ function, and the entropy becomes zero.
These arguments hold similarly
for the case of higher dimensions or the case of 
density functional theory. 

For higher
temperatures $T$, the forces and the 
derivative of
the total energy deviate stronger, and a suggestion was made to remedy this,
in order to obtain the derivative of the total energy, and not of the
free energy \cite{Wagner1998}.

\section{Examples}
\label{examples}

In the following, some examples demonstrate the accuracy of the
gradients. The calculations were done with the present CRYSTAL09 release
\cite{Manual09,CRYSTALbuch}. The examples aim at documenting the
accuracy of the gradient, by comparing the analytical and numerical
gradient, at the level of Hartree-Fock and density functional theory,
for the gradient with respect to the cell parameter, and with
respect to the nuclear position. 

First, for Cu bulk, the analytical and numerical 
gradient with respect to the cell parameter are compared in table
\ref{Cubulk}. This is done on the Hartree-Fock and density functional
level. The basis sets from reference \cite{ClCu111paper} were used.
A $\vec k$-point mesh with 16 $\times$ 16 $\times$ 16 points was used.
Smearing temperatures in the range from 0.001 $E_h$ to 0.05 $E_h$ were chosen.
Technically, in the input,
a hybrid functional consisting of nothing but 100\% Fock exchange was defined,
in order to perform the Hartree-Fock calculation at finite temperature.
When comparing numerical and analytical derivatives, then
the obtained accuracy for the derivative of the
free energy $-\frac{\partial F}{\partial a}$ is
similar to the one for insulators, see \cite{Doll2001IJQC,Doll2001CPC,Doll2004TCA,Doll2006TCA,Doll2010}.
Note that in addition, the numerical noise is in general larger
in the case of metals, and therefore, also the energies and their
numerical derivatives carry larger noise.
The agreement between analytical and numerical derivative of the
free energy is similar for all smearing temperatures. 

\begin{figure}
\vspace{0cm}
\caption{Left: Free energy for Cu bulk; crosses refer to computed points, the
full line is a fit through the points.
Right: Analytical (crosses) and numerical derivative (full line)
with respect to the cell parameter for Cu bulk. The numerical derivative
is obtained as a derivative of the fit of the energy expression in the
left figure. A smearing temperature of 0.001 $E_h$ was applied.}

\vspace{1cm}

\hspace{-1cm}
\includegraphics[width=7cm,angle=0]{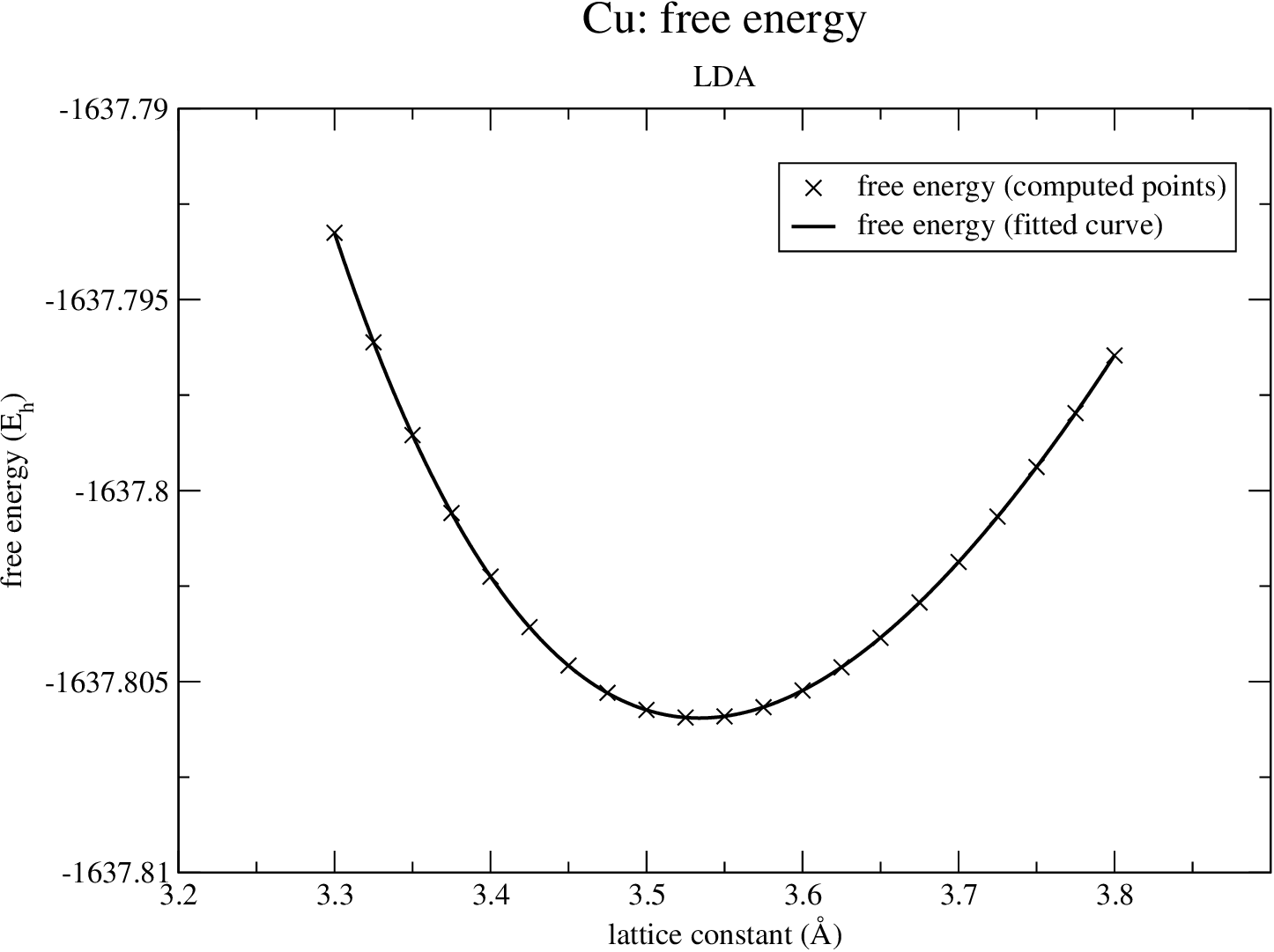}
\hspace{1cm}
\includegraphics[width=7cm,angle=0]{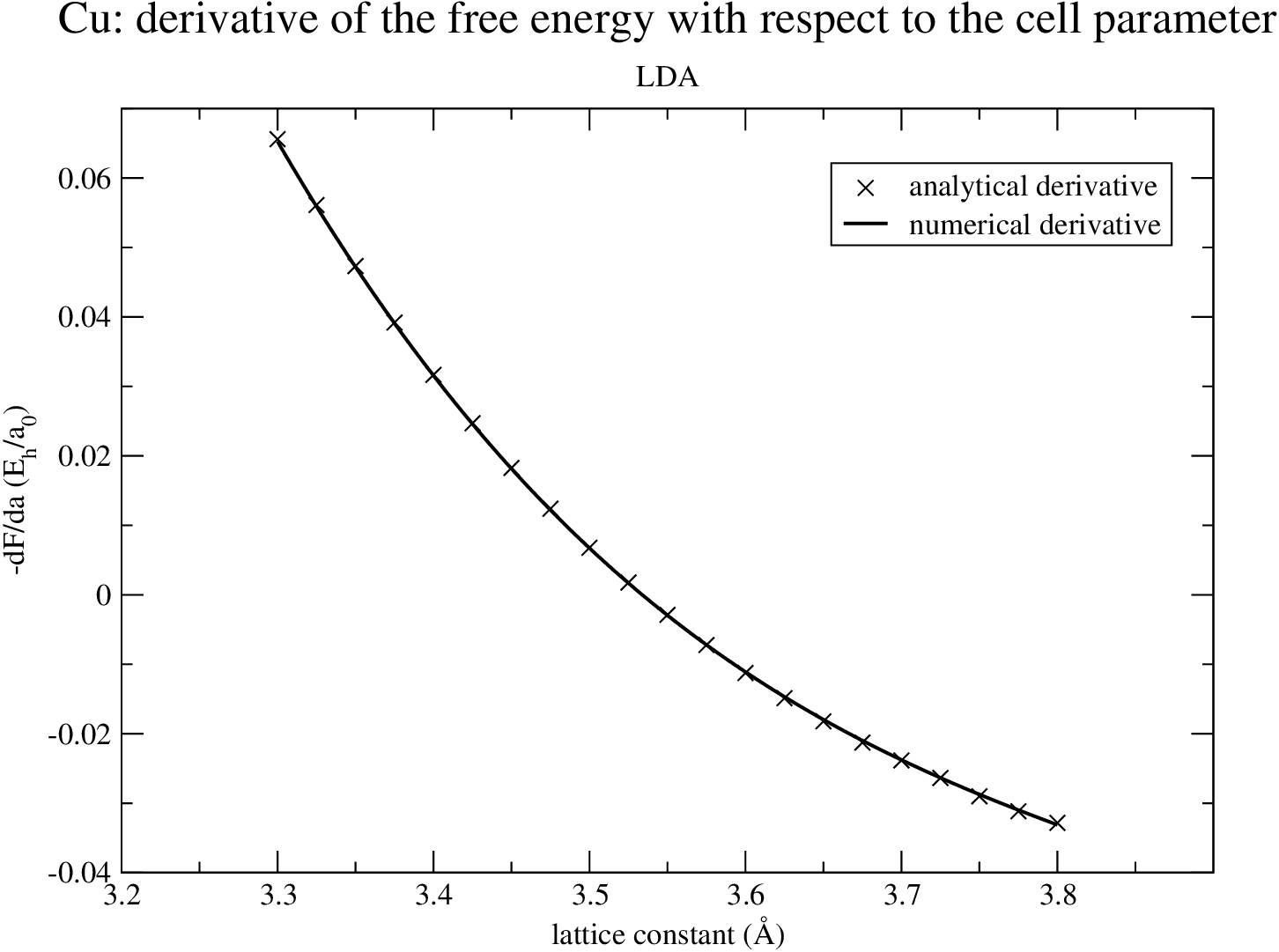}
\label{Cubulkgradientenbild}
\end{figure}

The derivative of the
energy with respect to the cell parameter agrees reasonably well
at low temperatures, but deviates strongly at high smearing temperatures,
as expected, as the energy and the free energy deviate more and more
at higher temperature. The free energy and its derivative with respect to
the cell parameter are also visualized in figure \ref{Cubulkgradientenbild},
where a smearing temperature of 0.001 $E_h$ was employed.
Again, the agreement between numerical and analytical derivative
is very good.

\begin{table}
\begin{center}
\caption{\label{Cubulk}The derivative of the total energy and the
free energy, in hartree/bohr
($E_h/a_0$), with
respect to the cell parameter $a$, analytical and numerical, on the
Hartree-Fock and density functional (LDA) level.}
\vspace{5mm}
\begin{tabular}{cccccc}
smearing 
temperature &  $-\frac{\partial E}{\partial a}$ (numerical) & $-\frac{\partial
  F}{\partial a}$ (numerical)
& $-\frac{\partial F}{\partial a}$ (analytical)\\
($E_h$) & ($\frac{E_h}{a_0}$) & ($\frac{E_h}{a_0}$) & ($\frac{E_h}{a_0}$) \\
\hline\hline
\multicolumn{4}{c}{Hartree-Fock (at $a=5$ \AA)}\\ \\
0.001 & -0.0316 & -0.0316 & -0.0314 \\
0.01 & -0.0317 & -0.0315 & -0.0313 \\
0.03 & -0.0328 & -0.0305 & -0.0303 \\
0.05 & -0.0352 & -0.0276 & -0.0280 \\
\\
\multicolumn{4}{c}{LDA (at $a=3.4$ \AA)}\\ \\
0.001 & 0.0315 & 0.0315 & 0.0317 \\
0.01 & 0.0310 & 0.0319 & 0.0320 \\
0.03 & 0.0212 & 0.0390 & 0.0393 \\
0.05 & 0.0098 & 0.0540 & 0.0542 \\
\hline\hline
\end{tabular}           
\end{center}
\end{table}

\begin{figure}
\vspace{0cm}
\caption{Left: Free energy for Cl/Cu(111); 
crosses refer to computed points, the
full line is a fit through the points.
Right: Analytical (crosses) and numerical derivative (full line)
with respect to the z-position of the Cl atom. The numerical derivative
is obtained as a derivative of the fit of the energy expression in the
left figure. A smearing temperature of 0.001 $E_h$ was applied.}

\vspace{1cm}

\hspace{-1cm}
\includegraphics[width=7cm,angle=0]{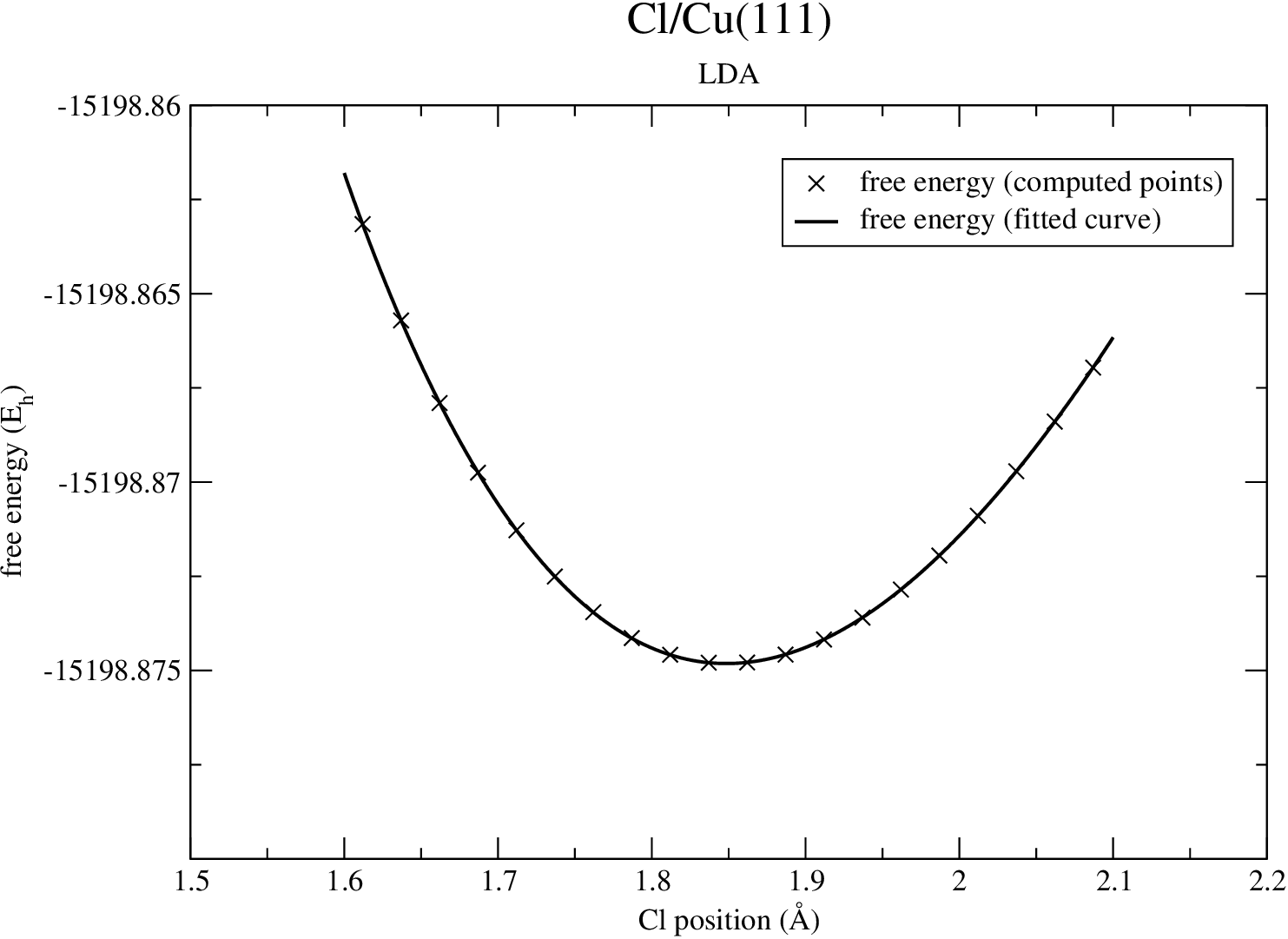}
\hspace{1cm}
\includegraphics[width=7cm,angle=0]{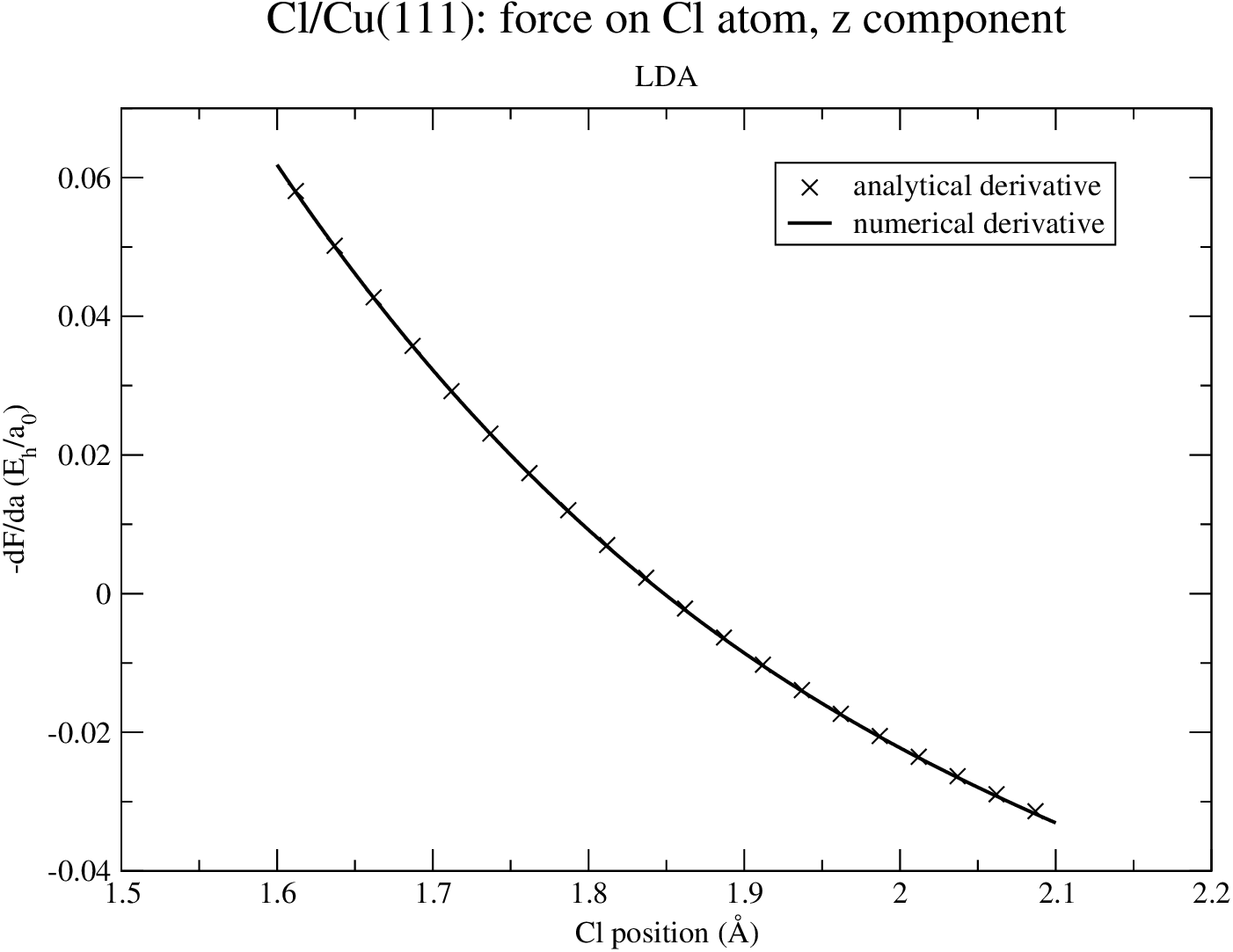}
\label{ClCu111gradientenbild}
\end{figure}

As an example for the gradient with respect to nuclear positions,
the adsorbate system 
Cu(111)$(\protect\sqrt{3} \times \protect\sqrt{3})$R30$^\circ$-Cl
is considered, with chlorine sitting on the hcp (hexagonal close packed)
site. The basis sets are as in \cite{ClCu111paper},
and 16 $\times$ 16 $\vec k$-points together with a smearing temperature of 
0.001 $E_h$ is used. The free energy and its
derivative with respect to the z-component of the
chlorine atom are computed analytically and numerically. The results
are visualized in figure \ref{ClCu111gradientenbild}, and the
numerical and analytical derivatives agree well. The computed equilibrium
position corresponds to a hight of 1.85 \AA \ above the topmost Cu layer, in
reasonable agreement with the earlier calculation \cite{ClCu111paper}:
in the earlier calculation, a generalized gradient functional had been employed
and a hight of 1.90 \AA \ had been obtained. 
The present calculation gives a
slightly shorter bond length which is a usual feature of the local
density approximation (LDA), as
compared to gradient corrected functionals. Note that no gradients
had been used in the earlier work \cite{ClCu111paper}, and the
geometry had been determined by iteratively optimizing the various
geometrical parameters, by employing the total energy only.

\section{Conclusion}
Derivatives of the total and free energy of periodic systems
with respect to geometrical parameters were studied theoretically, 
in the case of metallic systems. In the case of metals, numerical integration
is often facilitated by introducing an artificial temperature and by
an occupancy according to e.g. the Fermi function.
At zero temperature,
the theory of the derivatives does not require an additional term
compared to the case of insulators. 
At finite temperature, when the occupancy is according to the Fermi
function, then a similar expression for the derivative can
be employed, which is however only consistent with the free energy.
Therefore, numerical derivatives of the free energy agree reasonably well
with analytical derivatives, and consequently, numerical derivatives 
of the total
energy deviate more and more with increasing temperature.
This holds for the case of Hartree-Fock or density functional theory.
Numerical examples demonstrate the accuracy which is achieved with
the implementation in the CRYSTAL code.





\bibliographystyle{elsarticle-num}
\bibliography{<your-bib-database>}

\begin{thebibliography}{00}


\bibitem{CRYSTALbuch} C. Pisani, R. Dovesi, and C. Roetti,
Hartree-Fock Ab Initio Treatment of Crystalline Systems, 
Lecture Notes in Chemistry Vol. 48, Springer, Heidelberg, 1988.
\bibitem{EvarestovBuch}
R. A. Evarestov, Quantum Chemistry of Solids, 
Springer Series in Solid-State Sciences,
Vol. 153, Springer, Berlin, Heidelberg, New York, 2007.
\bibitem{Pulay} P. Pulay, Mol. Phys. 17 (1969) 197.
\bibitem{Bratoz} S. Brato\u{z}, in 
{\it Calcul des fonctions d'onde mol{\'e}culaire},
Colloq. Int. C. N. R. S. 82 (1958) 287.
\bibitem{Schlegel2000} H. B. Schlegel, Theor. Chim. Acta 103 (2000)
294.
\bibitem{Teramae1983} H. Teramae, T. Yamabe, C. Satoko and A. Imamura,
Chem. Phys. Lett.  101 (1983) 149.
\bibitem{Teramae1984}
H. Teramae, T. Yamabe and A. Imamura, J. Chem. Phys.  81 (1984) 3564.
\bibitem{Feibelman1991a} P. J. Feibelman, Phys. Rev. B  35 (1987) 2626.
\bibitem{Feibelman1991b} P. J. Feibelman, Phys. Rev. B  44 (1991) 3916.
\bibitem{HirataIwata1997} S. Hirata and S. Iwata, J. Chem. Phys.  107 (1997)
10075.
\bibitem{Sun1998} J.-Q. Sun and R. J. Bartlett, J. Chem. Phys. 109 (1998)
4209.
\bibitem{Jacquemin1999a} D. Jacquemin, J.-M. Andr{\'e} and B. Champagne,
J. Chem. Phys.  111 (1999) 5306. 
\bibitem{Jacquemin1999b} D. Jacquemin, J.-M. Andr{\'e} and B. Champagne,
J. Chem. Phys.  111(1999) 5324 .
\bibitem{KudinScuseria2000} K. N. Kudin and G. E. Scuseria, Phys. Rev. B
 61 (2000) 5141.
\bibitem{Doll2001IJQC} K. Doll, V. R. Saunders, N. M. Harrison,
Int. J. Quantum Chem.  82 (2001) 1.
\bibitem{Doll2001CPC} K. Doll, Comp. Phys. Comm.  137 (2001) 74.
\bibitem{Doll2004TCA} K. Doll, R. Dovesi and R. Orlando,
Theor. Chem. Acc.  112 (2004) 394.
\bibitem{Doll2006TCA} K. Doll, R. Dovesi and R. Orlando,
Theor. Chem. Acc.  115 (2006) 354.
\bibitem{Doll2010} K. Doll, Mol. Phys. 108 (2010) 223.
\bibitem{Tobita2003} M. Tobita, S. Hirata, and R. J. Bartlett, J. Chem.
Phys. 118 (2003) 5776.
\bibitem{Weber2006} V. Weber, C. Daul, and M. Challacombe, J. Chem.
Phys.  124 (2006) 214105.
\bibitem{AshcroftMermin} N. W. Ashcroft and N. D. Mermin,
Solid State Physics, Saunders, Philadelphia (1976).
\bibitem{Goedecker1999} S. Goedecker, Rev. Mod. Phys.  71 (1999) 1085.
\bibitem{Heyd2003} J. Heyd, G. E. Scuseria, and M. Ernzerhof, J. Chem. 
Phys.  118 (2003) 8207.
\bibitem{Pierobuchartikel} K. Doll,
{\it Ab initio calculations with a Gaussian
basis set for metallic surfaces and the adsorption thereon}, in
Quantum Chemical Calculations of Surfaces and Interfaces of Materials, 
edited by Vladimir Basiuk and Piero Ugliengo,
American Scientific Publishers, 2009, pp. 41-53.
\bibitem{Kertesz1984} M. Kertesz, Chem. Phys. Lett.  106 (1984) 443.
\bibitem{SzaboOstlund}
A. Szabo and N. S. Ostlund,
Modern Quantum Chemistry, MacGraw-Hill, New York, 1989.
\bibitem{VicCoulomb} V. R. Saunders, C. Freyria-Fava, R. Dovesi, L. Salasco,
and C. Roetti, Mol. Phys.  77 (1992) 629.
\bibitem{Vic1994} V. R. Saunders, C. Freyria-Fava, R. Dovesi, and C. Roetti,
Comp. Phys. Comm.  84 (1994) 156.
\bibitem{DelRe} G. Del Re, J. Ladik, G. Bicz\'{o}, Phys. Rev.  155
(1967) 997.
\bibitem{Andre} J. M. Andr\'e, J. Chem. Phys.  50  (1969) 1536.
\bibitem{Mermin1965} N. D. Mermin, Phys. Rev.  137 (1965) A1441.
\bibitem{FuHo1983} C.-L. Fu and K.-M. Ho, Phys. Rev. B  28 (1983) 5480.
\bibitem{Ho1992} K. H. Ho, C. Els\"asser, C. T. Chan, and M. F\"ahnle,
J. Phys.: Condens. Matt.  4 (1992) 5189.
\bibitem{Elsaesser1994} C. Els\"asser, M. F\"ahnle, C. T. Chan and
K. M. Ho, Phys. Rev. B  49 (1994) 13975.
\bibitem{Springborg1998} M. Springborg, R. C. Albers, and K. Schmidt,
Phys. Rev. B 57 (1998) 1427.
\bibitem{Grabowski2007} B. Grabowski, T. Hickel, and J. Neugebauer,
Phys. Rev. B 76 (2007) 024309.
\bibitem{Gillan1989} M. J. Gillan, J. Phys.: Condens. Matt.  1  (1989) 689. 
\bibitem{Weinert1992} M. Weinert and J. W. Davenport, 
Phys. Rev. B  45 (1992) 13709.
\bibitem{Wentzcovitch1992} R. M. Wentzcovitch, J. L. Martins, and
P. B. Allen, Phys. Rev. B  45 (1992) 11372.
\bibitem{Warren1996} R. W. Warren and B. I. Dunlap, Chem. Phys. Lett.
 262 (1996) 384.
\bibitem{Wagner1998} F. Wagner, Th. Laloyaux, and M. Scheffler, 
Phys. Rev. B  57 (1998) 2102.
\bibitem{Manual09} 
R. Dovesi, V. R. Saunders, C. Roetti, R. Orlando, C. M. Zicovich-Wilson,
F. Pascale, B. Civalleri, K. Doll, N. M. Harrison, I. J. Bush, Ph. D'Arco,
M. Llunell, CRYSTAL2009, University of Torino, Torino, 2009.
\bibitem{ClCu111paper} K. Doll and N. M. Harrison, 
Chem. Phys. Lett.  317 (2000) 282.
\end{thebibliography}



\newpage

\newpage
\clearpage

\end{document}